\documentclass[english,prd,nofootinbib,preprintnumbers,twocolumn,showpacs]{revtex4}
\usepackage[english]{babel}
\usepackage{amsmath}
\usepackage{amsfonts}
\usepackage{amssymb}
\usepackage{epsfig}
\usepackage{graphics,psfrag,rotating}
\usepackage{graphicx}
\usepackage{dcolumn}
\usepackage{bm}
\usepackage{epstopdf}
\usepackage{color}
\usepackage[usenames,dvipsnames,svgnames]{xcolor}
\usepackage[colorlinks=true,
            linkcolor=red,
            urlcolor=gray,
            citecolor=blue]{hyperref}
  \usepackage{hyperref}
\usepackage[T1]{fontenc}
\usepackage{multirow}
\usepackage{float}

\usepackage{subfigure}

\usepackage{enumitem}

\newcommand{\nn}{\nonumber\\}

\def\3nab{\tilde{\nabla}}

\def\be {\begin{equation}}
\def\ee {\end{equation}}
\def\ba {\begin{align}}
\def\ea {\end{align}}

\def\bc {\begin{center}}
\def\ec {\end{center}}
\def\case#1/#2{\frac{#1}{#2}}

\newcommand{\bver}{\begin{verbatim}}
\newcommand{\ever}{\end{verbatim}}

\newcommand{\bea}{\begin{eqnarray}}
\newcommand{\eea}{\end{eqnarray}}
\newcommand{\beaa}{\begin{eqnarray*}}
\newcommand{\eeaa}{\end{eqnarray*}}

\def\case#1/#2{\textstyle\frac{#1}{#2}}

\begin{document}

\title{Is cosmography a useful tool for testing cosmology?}

\author{
Vinicius C. Busti$^{a}\,\footnote{vcbusti [at] astro.iag.usp.br}$
\'Alvaro de la Cruz-Dombriz$^{a,b}\, \footnote{ alvaro.delacruzdombriz [at] uct.ac.za}$ 
Peter K. S. Dunsby$^{a,c}\,\footnote{ peter.dunsby [at] uct.ac.za}$, 
Diego S\'aez-G\'omez$^{a,d}\,\footnote{dsgomez [at] fc.ul.pt}$
}
\affiliation{
$^{a}$ Astrophysics, Cosmology and Gravity Centre (ACGC), Department of Mathematics and Applied Mathematics, University of Cape Town, Rondebosch 7701, Cape Town, South Africa.\\
$^{b}$ Departamento de F\'{\i}sica Te\'orica I, Ciudad Universitaria, Universidad Complutense de Madrid, E-28040 Madrid, Spain.\\
$^{c}$ South African Astronomical Observatory,  Observatory 7925, Cape Town, South Africa\\
$^{d}$ Departamento de F\'isica \& Instituto de Astrof\'isica e Ci\^encias do Espa\c{c}o,\\ 
Faculdade de Ci\^encias da Universidade de Lisboa, Edif\'icio C8, Campo Grande, P-1749-016
Lisbon, Portugal}

\pacs{98.80.Jk,\, 04.20.Cv,\, 04.50.Kd,\,95.36.+x}

\begin{abstract} 

Model-independent methods in cosmology has become an essential tool in order to deal with an increasing number of theoretical alternatives for explaining the late-time acceleration of the Universe. In principle, this provides a way of testing the Cosmological Concordance (or $\Lambda$CDM) model under different assumptions and ruling out whole classes of competing theories. One such model-independent method is the so-called {\em cosmographic approach}, which relies only on the homogeneity and isotropy of the Universe on large scales. We show that this method suffers from many shortcomings, providing biased results depending on the auxiliary variable used in the series expansion and is unable to rule out models or adequately reconstruct theories with higher-order derivatives in either the gravitational or matter sector. Consequently, in its present form, this method seems unable to provide reliable or useful results for cosmological applications.

\end{abstract} 


\maketitle

\section{Introduction}

The phenomenon driving the acceleration of the Universe remains one of the biggest puzzles in fundamental physics today. In order to deal with this problem many models have been proposed, ranging from assuming an effective perfect fluid with negative pressure, dubbed dark energy, modifications 
of General Relativity (GR) on large scales or even breaking the Copernican Principle (see \cite{Copeland:2006wr} for reviews). As the number of possibilities increases, a useful tool is to use frameworks, which are able to encompass a large class of models or theories and where observational scrutiny 
is able to rule out whole classes of possibilities without having to analyse each one on a case-by-case basis.  Such model-independent methods have been applied to infer the dark energy equation of state and to reconstruct classes of dark energy theories. Among them, the {\em cosmographic approach} \cite{Weinberg-Harrison}, 
which relies on the Copernican Principle (leading to the Friedmann-Lema\^{i}tre-Robertson-Walker (FLRW) metric), and the expression of the scale factor as a function of an auxiliary variable - either time or redshift for instance -  have been used to look for deviations in the standard $\Lambda$CDM model or to reconstruct models for dark energy and the action for modified gravity \cite{Bernstein:2003es,Visser,Aviles_2012,Bamba:2012cp,Capozziello_PRD_fR,Capozziello_Ruth_PRD}.

In this paper, we discuss several limitations of cosmographic analyses to constrain various models as well as their ability to reconstruct theories from cosmological data. 
\section{Cosmography approach}

The cosmography approach starts by defining the cosmographic functions, 
\begin{equation}
H=\frac{\dot{a}}{a}\ ,\, q=-\frac{\ddot{a}}{aH^2}\ , \, j=\frac{a^{(3)}}{aH^3}\ , \,  s=\frac{a^{(4)}}{aH^4}\ , \,  l=\frac{a^{(5)}}{aH^5}\ , ...\ 
\label{parameters}
\end{equation}
where $a$ is the scale factor, the dots represent cosmic time derivatives and $a^{(n)}$ holds for the $n^{\rm th}$ time derivative of $a$. Evaluated today,
those functions define the cosmographic parameters $H_0$, $q_0$, $j_0$, $s_0$ and $l_0$, which are the Hubble constant, the deceleration parameter,
the jerk parameter, the snap parameter and the lerk parameter, respectively. Then, by using the definition of the redshift $1+z=\frac{1}{a}$, the Hubble parameter $H$ can be expanded in powers of the redshift around $z=0$ as $H=H_0+H_{z0}z+\frac{H_{zz0}}{2}z^2+...$, which converges for $|z|<1$ and where the Hubble parameter derivatives, denoted by a $z$ subscript, are evaluated today (subscript $0$). They can be expressed in terms of the cosmographic parameters as follows (throughout this paper, let us assume that the curvature density parameter $\Omega_k$ is negligible):
\bea
H_{z0}/H_0&=&1+q_0\ ,\quad H_{zz0}/H_0=-q_0^2+j_0\ , \nn
H_{3z0}/H_0&=&3q_0^2(1+q_0)-j_0(3+4q_0)-s_0\ , \nn
H_{4z0}/H_0&=&-3q_0^2(4+8q_0+5q_0^2)+j_0\left(12+32q_0\right. \nn
&&\left.+25q_0^2-4j_0\right)+s_0(8+7q_0)+l_0\ .
\label{HubbleDerivsZ}
\eea
In principle the parameters (\ref{HubbleDerivsZ}) can be fitted with observational data, leading to model-independent constraints. As an alternative to the independent variable $z$, the above expansion may be expressed in terms of a more convenient variable that ensures the convergence of the series for the whole past history of the universe \cite{Visser}:
\begin{equation}
y = \frac{z}{1+z}.
\end{equation}
In order to illustrate the differences between the $z$ and $y$ redshift measures, let us use mock data generated from a fiducial spatially flat $\Lambda$CDM model, assuming the same 
redshift distribution as the Union 2.1 catalogue \cite{union2.1}, with errors of magnitude $\sigma_{\mu}=0.15$.  In this paper we restrict ourselves to the $y$ and $z$ parameters and supernova type data. Although many different observables
have been used to constrain cosmographic parameters, such as $H(z)$ data, baryon acoustic oscillations, gamma-ray bursts, angular distances to galaxy clusters \cite{cosm_constraints}, 
it is easier to 
understand what is going on with only one observable. Thus we leave combined probes analyses for future work.

\begin{table}[htbp]
\caption{Coverage test for $\bm{\theta_1}$ and $\bm{\theta_2}$. Refer to the bulk of the text for further details.}
\label{tables1}
\begin{center}
\begin{tabular}{@{}cccccccccccccc@{}}
\hline
&     &     &     &  $\bm{\theta_1}$   &      &     &     &      &     &  $\bm{\theta_2}$   &      &     \\ \hline 
& \vline    &   $y$   &   \vline  & \vline    &  $z$   &   \vline &   & \vline    &   $y$   &   \vline  & \vline    &   $z$   &   \vline  \\
\hline  & $1\sigma$ & $2\sigma$ &
 3$\sigma$ & $1\sigma$ & $2\sigma$ &
 3$\sigma$ &  & $1\sigma$ & $2\sigma$ &
 3$\sigma$ & $1\sigma$ & $2\sigma$ &
 3$\sigma$
\\ \hline
$q_0$ & 26 & 32 & 42 & 67   & 27 & 6 &  & 82 & 12 & 6 & 82   & 18 & 0  \\ 
$j_0$ & 10  &  45 & 45 & 64  & 29 & 7 &  & 93  &  5 & 2 & 88  & 12 & 0  \\ 
$s_0$ & 10  & 67 & 23 & 83  & 15 & 2 &  & 92  & 7 & 1 & 93  & 6 & 1 \\
$l_0$ & - & - & - & - & - & - &  & 100 & 0 & 0 & 100 & 0 & 0 \\
\hline
\end{tabular}
\end{center}
\end{table}
By generating 100 simulations based on this model, with $\Omega_m=0.3$ and $H_0=73.8$ km s$^{-1}$ Mpc$^{-1}$, we then constrain the cosmographic parameters (\ref{parameters}) by using both the $y$ and $z$ variables and then derive the posterior probability of the parameters. 
We have considered two sets of parameters: $\bm{\theta_1} = \{H_0,q_0,j_0,s_0\}$ and $\bm{\theta_2} = \{H_0,q_0,j_0,s_0,l_0\}$, where $H_0$ is analytically marginalized, since supernovae do not provide constraints over it. We have sampled from the posterior and obtained the constraints through a  
Monte Carlo Markov Chain (MCMC) using emcee \cite{emcee}. The constraints coming from the mock data sets allows us to infer how frequent the true values are within the 1, 2, and $3\sigma$ confidence regions, thereby providing a way to test whether there was any bias for a parameter for each of the two expansions performed ($y$ or $z$).

Table \ref{tables1} shows the number of times the true parameters were inside the confidence region bounds for the 
first set of parameters in the $y$ and $z$ parameterisations. We would expect the true value to lie within $1\sigma$, 
68\% of the times and $2\sigma$, 95\% of the times.  As shown, the $z$-parametrisation gives well-behaved coverage 
results. Only $s_0$ overestimates the errors. On the other hand, the $y$-parametrisation gives completely biased 
estimators, where only a few simulations are within $1\sigma$ and most of them lie at 3 or more $\sigma$. 
Interestingly enough, this happens despite the fact that the $y$-parametrisation has much larger errors than 
the $z$-parametrisation. 

\begin{figure*}
\begin{center}
\includegraphics[width=0.4\textwidth]{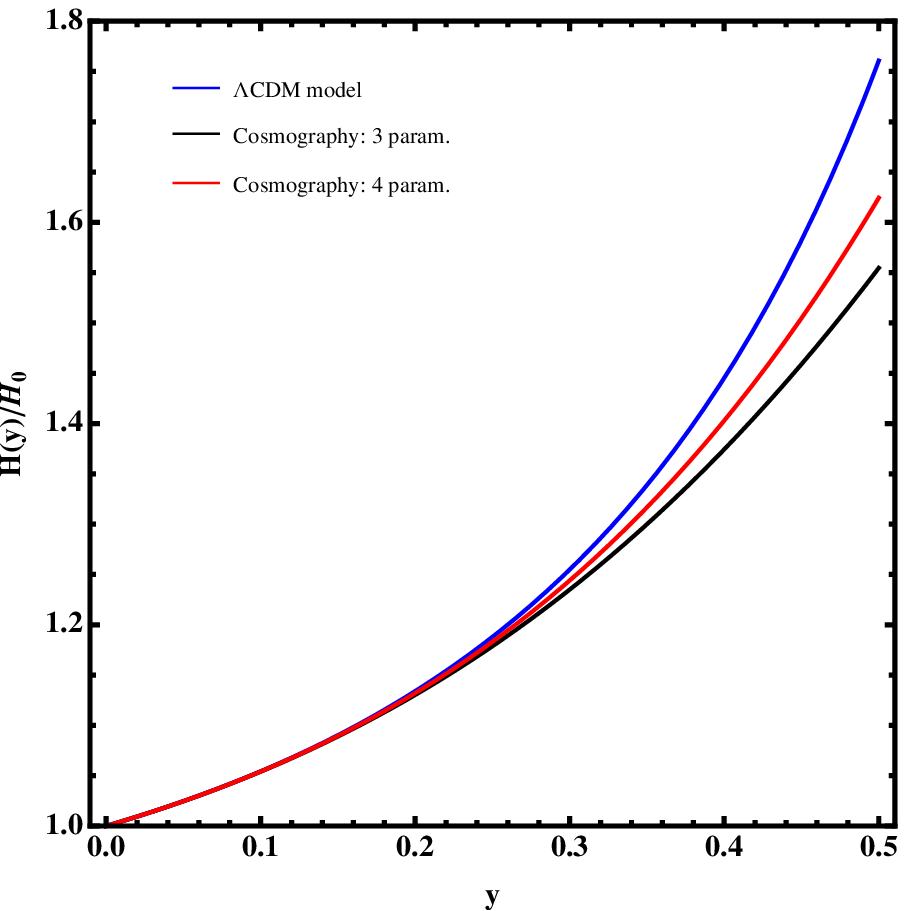}
\includegraphics[width=0.4\textwidth]{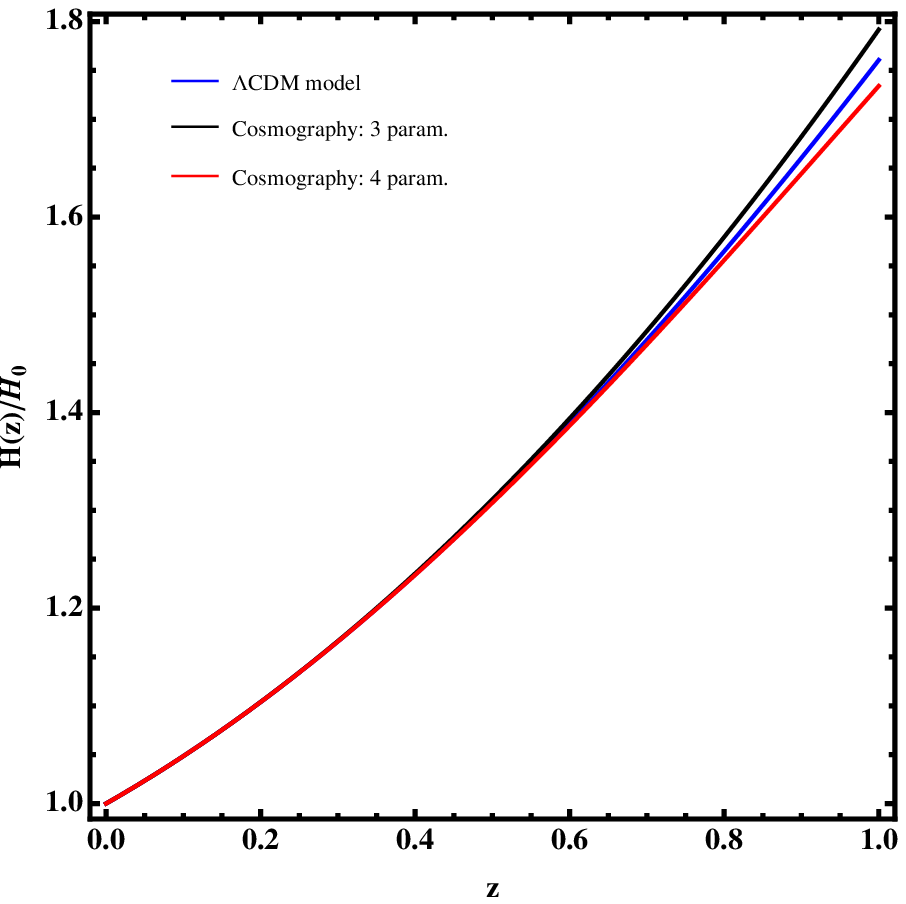}
\end{center}
\caption{Left panel: Hubble expansion rate for exact $\Lambda$CDM model with $\Omega_m=0.3$ and its corresponding cosmographic expansion with the $y$-parametrization with three
$\{q_0,j_0,s_0\}$ and four $\{q_0,j_0,s_0,l_0\}$
cosmographic parameters. Right panel: The same for the $z$-parametrization.}
\label{fig1a}
\end{figure*}

\begin{figure*}
\begin{center}
\includegraphics[width=0.485\textwidth]{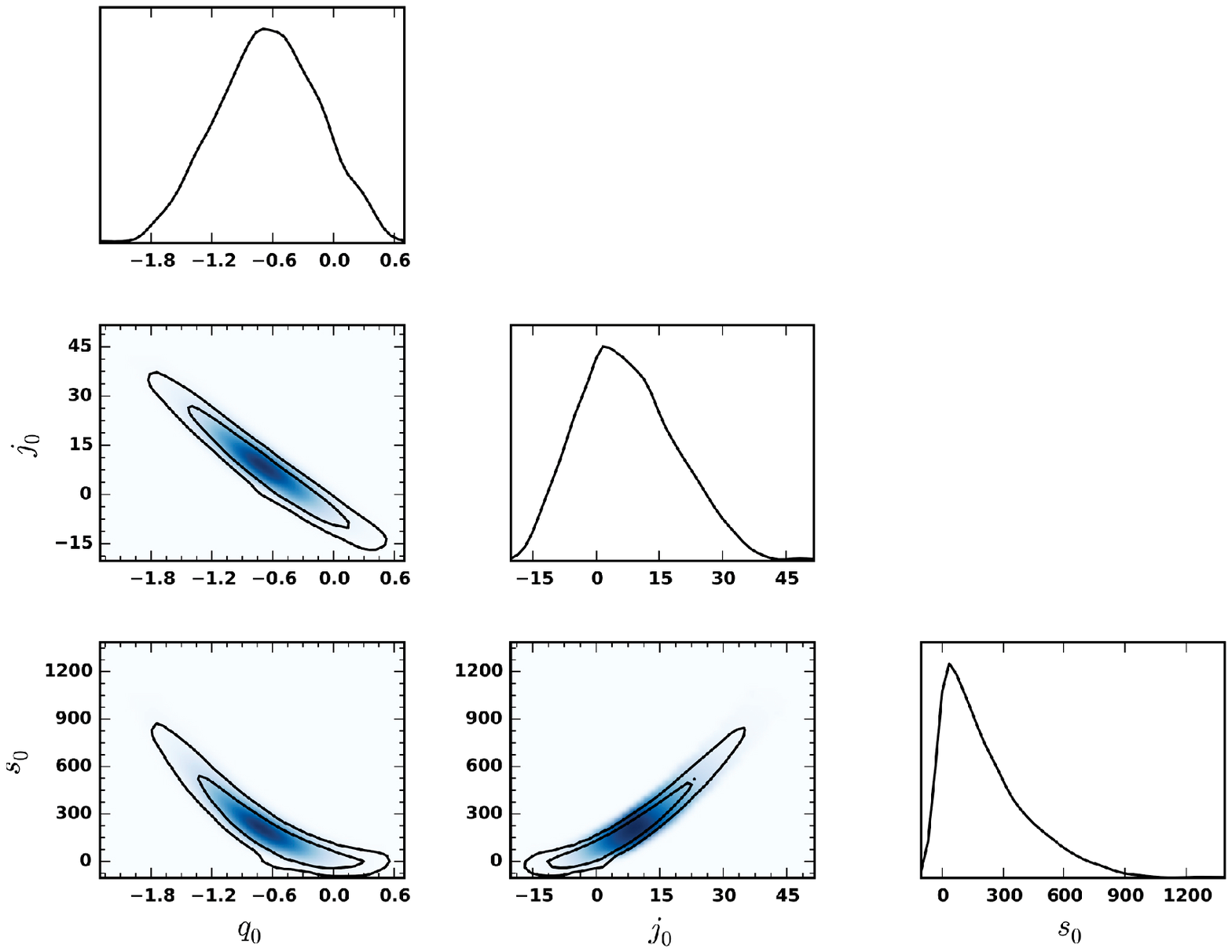}
\includegraphics[width=0.485\textwidth]{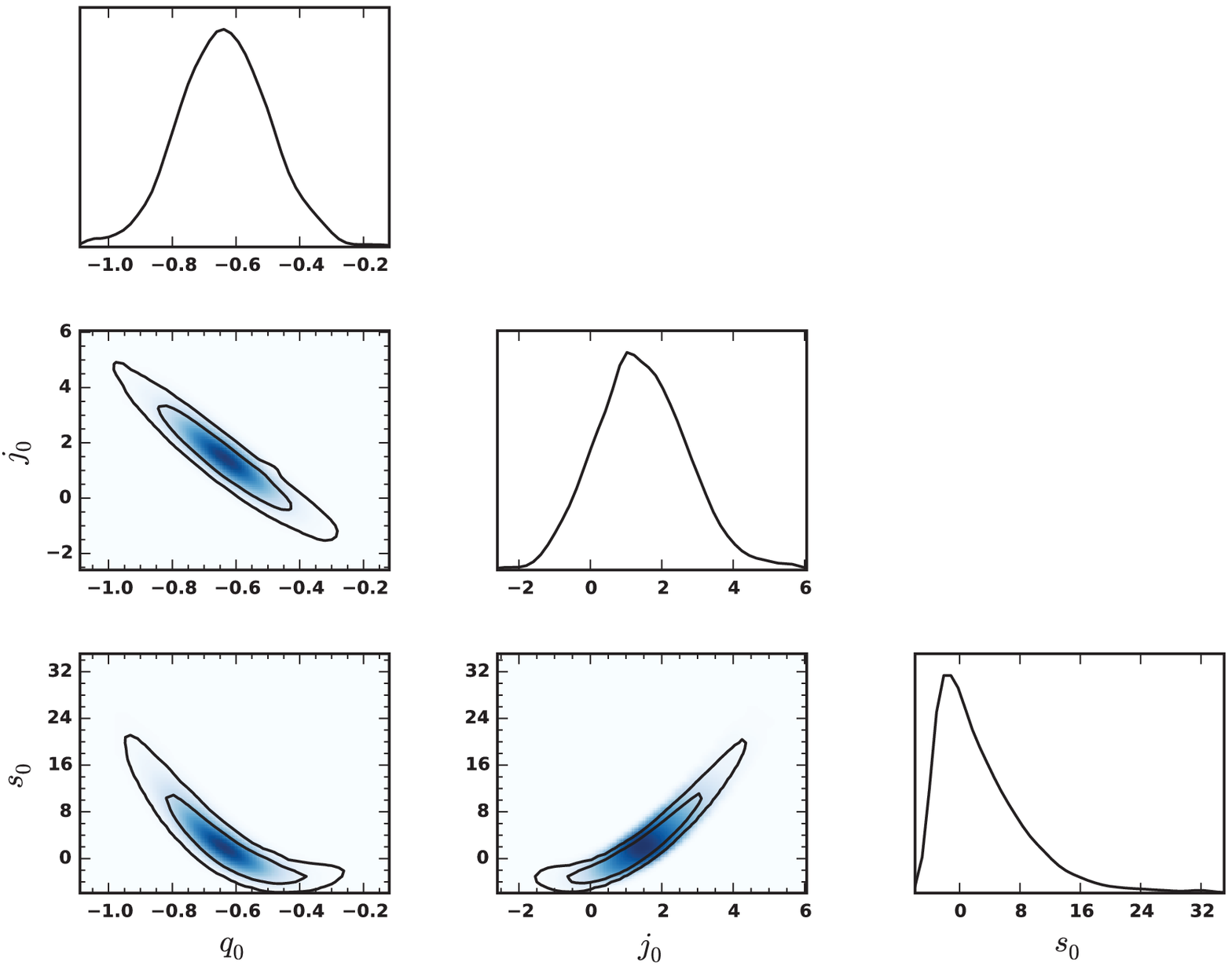}
\end{center}
\caption{Left panel: Constraints to the first set of cosmographic parameters with the $y$-parametrization. Right panel: The same for the $z$-parametrization.}
\label{fig1}
\end{figure*}

In the Figure \ref{fig1a} we have included a comparison between the exact $\Lambda$CDM Hubble parameter evolution and its corresponding expansion in terms of the cosmographic parameters following the ${\bf \theta_1}$ and $\bf{\theta_2}$ sets. For the $z$-redshift analysis, at $z\sim1$ the differences are around $0.6\%$ for both orders of the expansion whereas for the $y$-redshift analysis at a similar era $y\sim 0.5$ ($z\sim1$) the differences yield around $9\%$ while considering four parameters and even larger for three cosmographic parameters, showing the unexpected inadequacy of the $y$-redshift in a cosmographic analysis.  Figure \ref{fig1} then depicts a given realisation of our simulations, so the comparison of the size of the error can be made. 
The trends are as follows: the $y$-parametrisation provides much bigger errors, biasing
$q_0$ to smaller values and $j_0$ and $s_0$ to greater values than the true ones. 

In addition, while considering higher order terms in the expansion (see Table \ref{tables1}), the inclusion of a new parameter leads to an extreme overestimate of the errors for both variables, i.e., the values within $1\sigma$ for almost all realisations, where only $q_0$ is less affected. These results show that, despite $y$ might be preferable from a theoretical point of view, it is clearly not suitable to derive cosmological constraints, since the fittings are completely biased. 
One additional avenue that could be explored would be to derive a pivot redshift, around which the expansion is optimal. In theory, this could shed light on whether or not the behaviour we observe remains. Clearly, such a study depends strongly upon the redshift distribution of the sample under consideration and it is not within the scope of this paper. More to the point, this kind of study would be very helpful in order to understand how the redshift distribution of the sample influences the results, which might be important for some ongoing and future surveys as DES and LSST.
\section{Testing $\Lambda$CDM}
One of the predictions of the flat $\Lambda$CDM model regarding cosmography is that $j_0=1$. Consequently, we can easily check the consistency of the model by analyzing the posterior of this parameter. However, how the value of $j_0$ changes with the assumptions of the cosmographic analysis (for instance the order of the expansion) and how the constraints can be compared to the ones obtained when fitting the free parameters of the model, directly determine the usefulness of the test.

\begin{figure*}
\begin{center}
\includegraphics[width=0.45\textwidth]{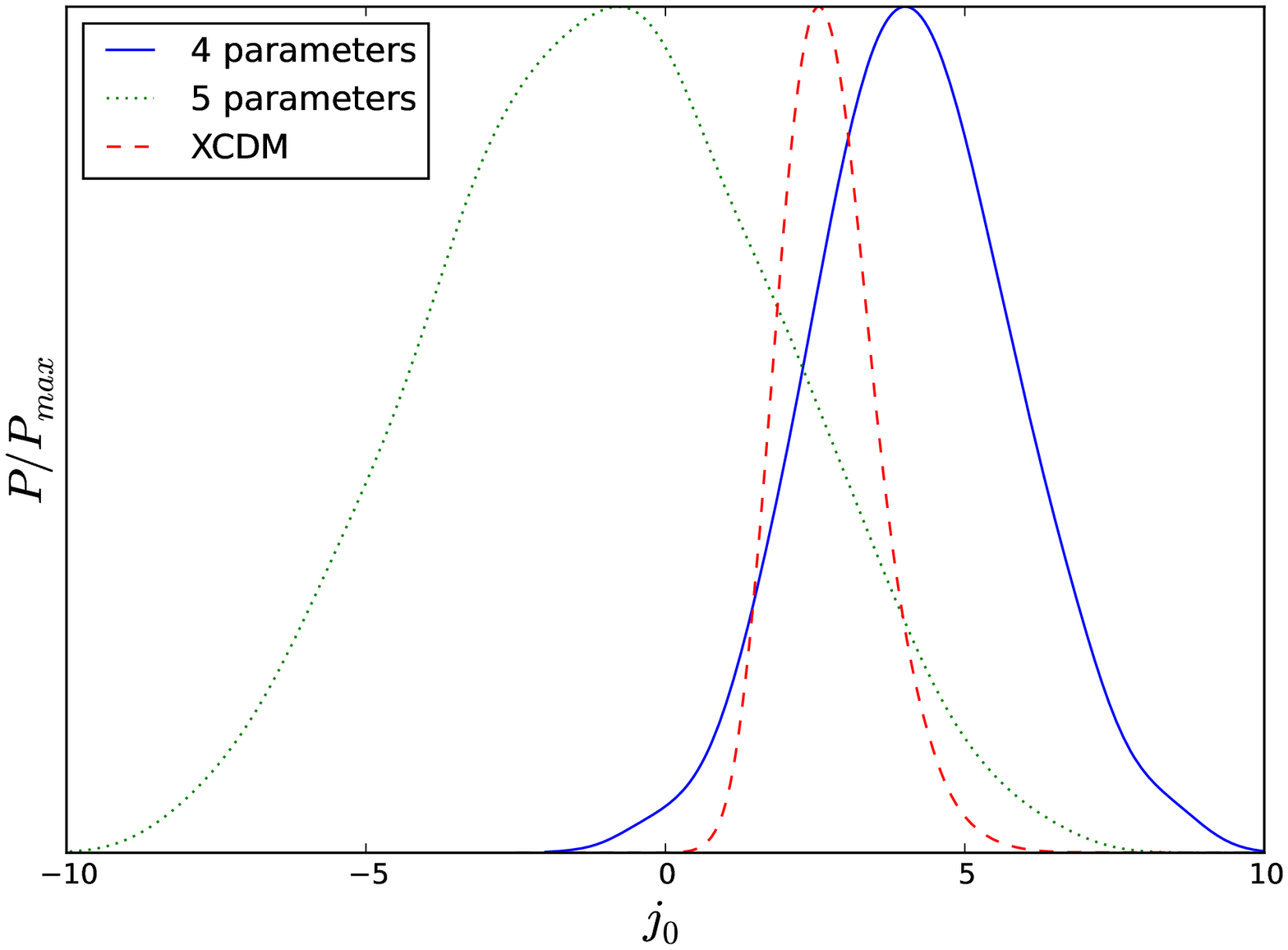}
\includegraphics[width=0.45\textwidth]{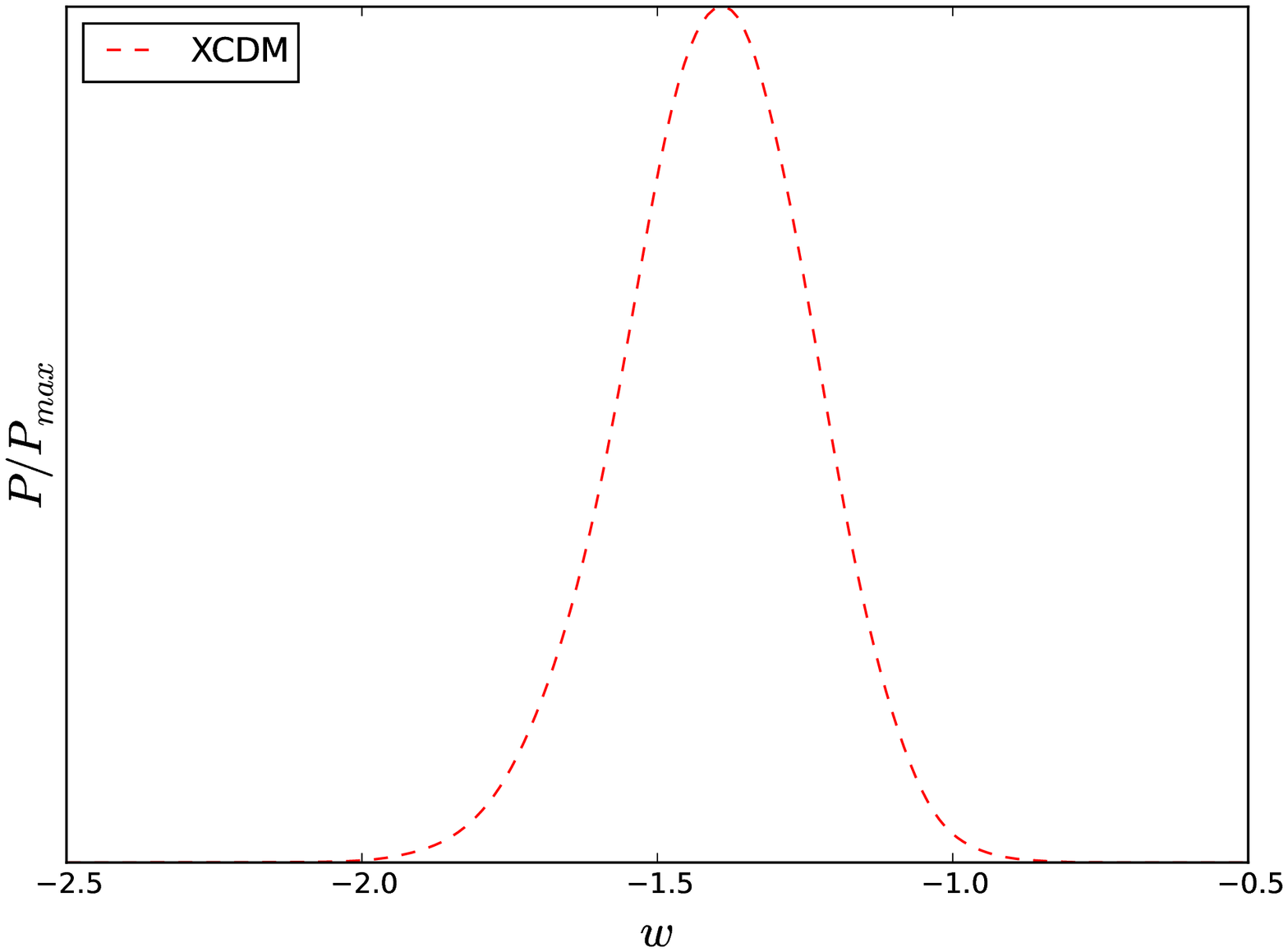}
\end{center}
\caption{Left panel: Posterior probability for $j_0$ considering 4 parameters $(\bm{\theta_1})$, 5 parameters $(\bm{\theta_1})$ and an XCDM model.
Right panel: Posterior probability for the dark energy equation of state parameter $w$. 
}
\label{fig2}
\end{figure*}

In order to address those questions we generated some mock realizations of data for a flat XCDM model, i.e., a model where the dark energy
equation of state is given by $w=p_{\rm X}/\rho_{\rm X}$, with $\Omega_m=0.3$ and $w=-1.3$, which gives $j_0=1.945$. Fig. \ref{fig2} (left and middle) displays the constraints of a typical realization considering the cosmographic expansion $\bm{\theta_1}$ (fourth order), $\bm{\theta_2}$ (fifth order) and the exact XCDM model, $H^2/H_0^2=\Omega_m (1+z)^3+(1-\Omega_m)(1+z)^{3(1+w)}$.
As shown in Fig.~\ref{fig2}, fitting the XCDM model directly provides smaller errors than the cosmographic approach, and it can spot deviations from $\Lambda$CDM with less effort. Moreover, the constraints on $w$ show a clear signature of $w \neq -1$. On the other hand, the order of the cosmographic expansion decisively affects the posterior constraints on the cosmographic parameters: there is some evidence of $j_0 \neq 1$ when considering $\bm{\theta_1}$ but such evidence disappears when $\bm{\theta_2}$ is assumed.

\section{Constraining dark energy models with cosmography analysis}

Recently it has been shown that cosmography can be used to reconstruct particular 
models for dark energy \cite{Bamba:2012cp}. However, as  demonstrated below, there are issues with this approach when it is applied to theories that contain higher derivatives, such as fourth-order gravity or Galileons. This is because the expansion of such models give extra free parameters apart from the usual cosmographic ones. We now compare several dark energy models beginning with one without higher derivatives:
\begin{equation}
\mathcal{S}=\int {\rm d}^4x^{}\sqrt{-g}\left[ \frac{1}{2}R 
 - \frac{1}{2} \omega (\phi)
\partial_{\mu} \phi \partial^{\mu }\phi -V(\phi )+\mathcal{L}_{m}\right]\ ,
\label{ST1}
\end{equation}
where $\mathcal{L}_m$ is the matter Lagrangian density,  $\omega(\phi)$ is the factor that renormalises the scalar field $\phi$ and $V(\phi)$ 
its potential. Note that for the sake of simplicity in the following we use units where $8\pi G=1$. Thus, the derivatives of the potential evaluated today, i.e., at redshift zero,  can be written as
\bea
&\frac{V_0}{H_0^2}&=2-q_0-\frac{3\Omega_m}{2}\ , \nn
&\frac{V_{z0}}{H_0^2}&=4+3q_0-j_0-\frac{9\Omega_m}{2}\ , \nn
&\frac{V_{2z0}}{H_0^2}&=4+8q_0+j_0(4+q_0)+s_0-9\Omega_m\ ,\\
&\frac{V_{3z0}}{H_0^2}&=j_0^2-l_0-q_0j_0(7+3q_0)-s_0(7+3q_0)-9\Omega_m\ .\nonumber
\label{ST7}
\eea
where field equations in a spatially-flat FLRW were used.

\begin{figure*}
\begin{center}
\includegraphics[width=0.45\textwidth]{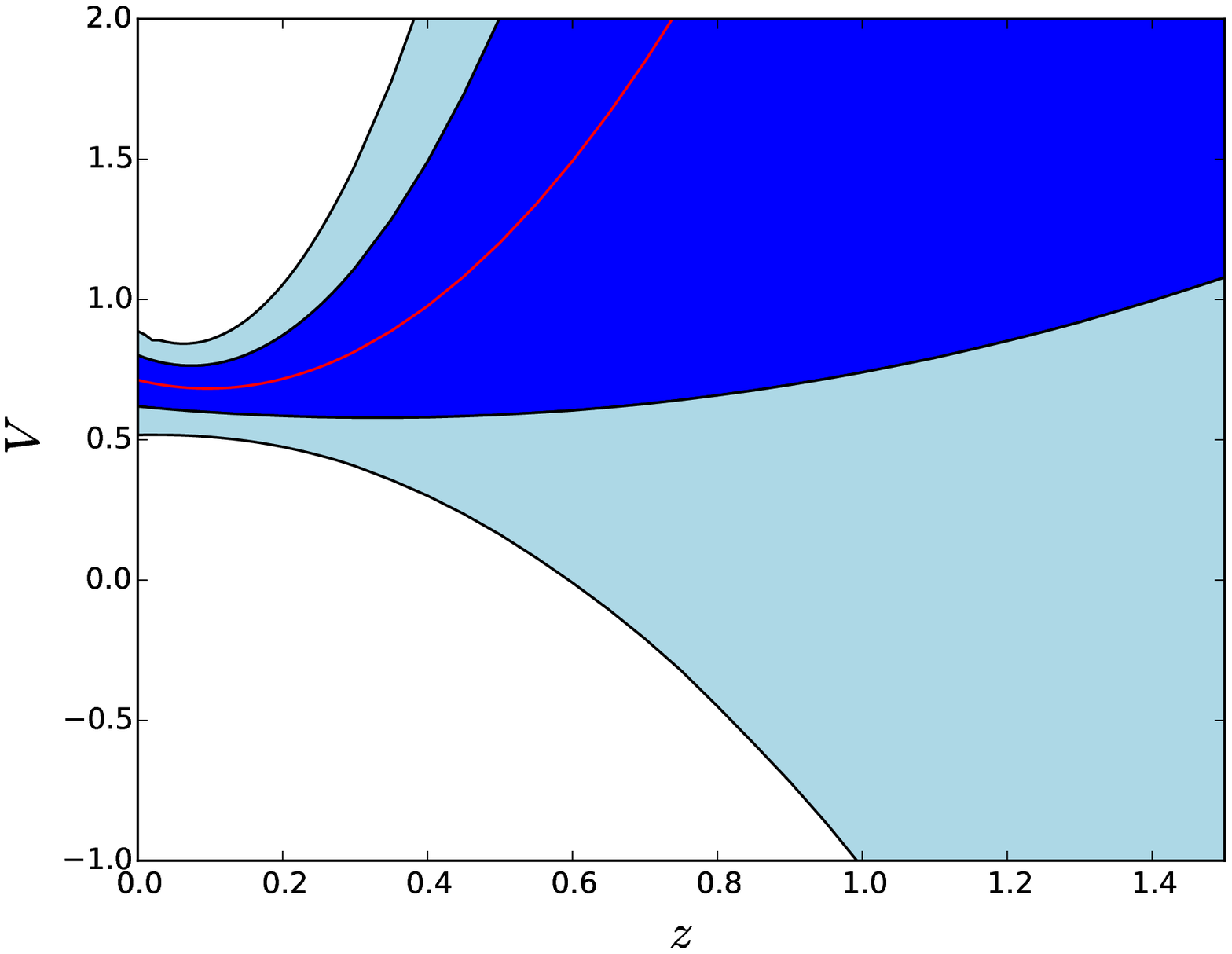}
\includegraphics[width=0.45\textwidth]{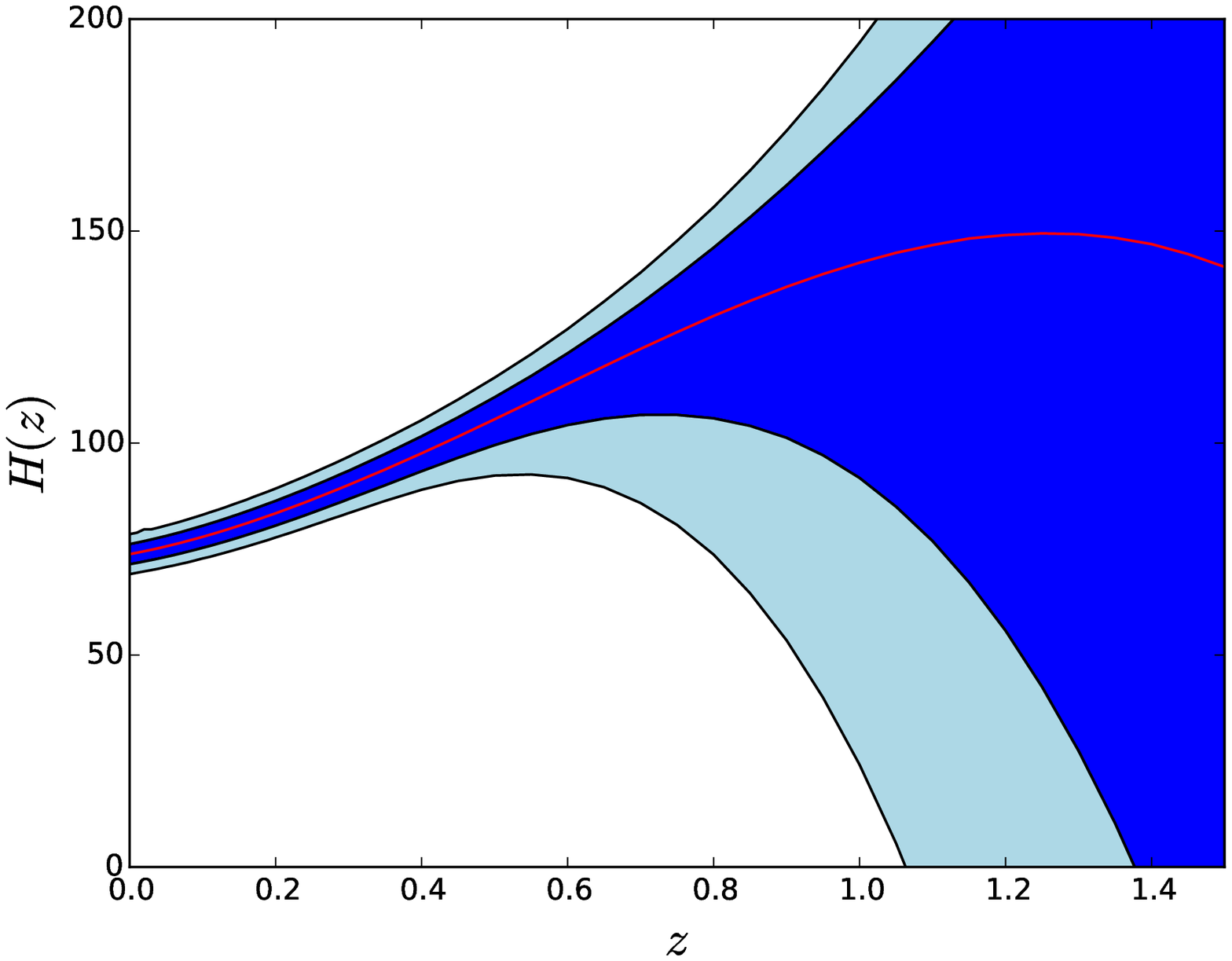}
\includegraphics[width=0.45\textwidth]{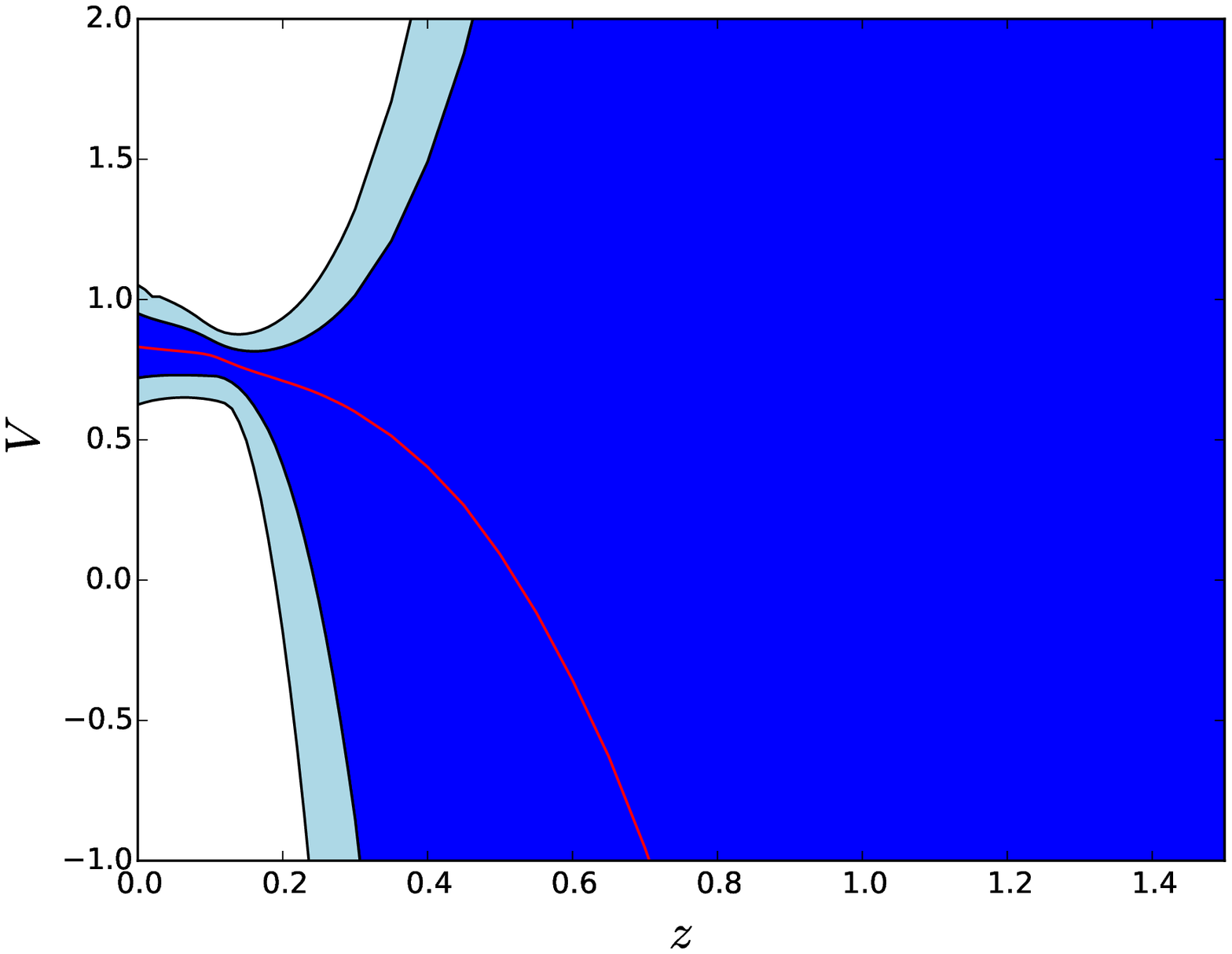}
\includegraphics[width=0.45\textwidth]{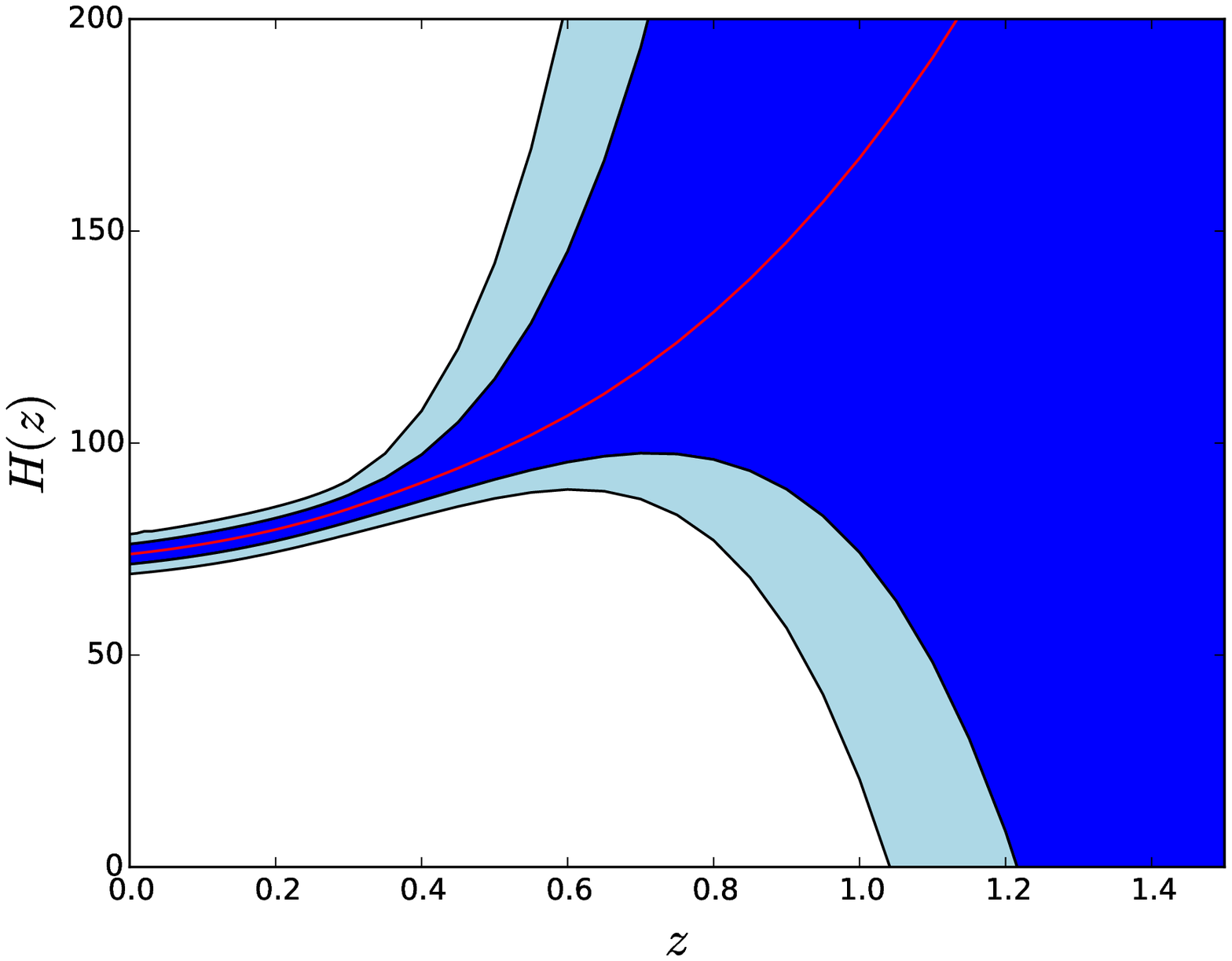}
\end{center}
\caption{Left upper panel: Scalar potential $V(\phi)$ evolution in terms of the redshift for four ($\bf{\theta_1}$)  parameters expansion. The light/dark blue regions represent the $1\sigma$/$2\sigma$ regions. Right upper panel: 
Hubble parameter for model (\ref{ST1}) as obtained from the fourth-order cosmographic parameters obtained in the MCMC analysis. Again light/dark blue regions represent the $1\sigma$/$2\sigma$ regions. Lower left and right panels are analogous to the upper counterparts for five ($\bf{\theta_2}$) parameters expansion.}
\label{fig_V}
\end{figure*}

Assuming the universe today is well approximated by a $\Lambda$CDM model, we can write $\Omega_m \approx 2/3(1+q_0)$. Thus, in this case there is a one-to-one correspondence between the derivatives of the potential and the cosmographic parameters. In this way, it is interesting to see how the method performs compared to other model-independent approaches. In Fig.~\ref{fig_V} (left), the evolution of the scalar potential $V(\phi)$ and the Hubble parameter (right) are shown for the $1\sigma$ and $2\sigma$ regions stopping at third and fourth orders for a generic realisation of a spatially flat $\Lambda$CDM model. This can be compared with a Gaussian process regression (see Fig. 1a in Ref.~\cite{nair}), where the errors are much smaller and no assumption regarding the model behaviour today was made. 
As can be seen, after $z \sim 0.5$ no useful constraints are derived for five parameters, while for four parameters good constraints are obtained up to $z \sim 1$.

\begin{figure*}
\begin{center}
\includegraphics[width=0.3\textwidth]{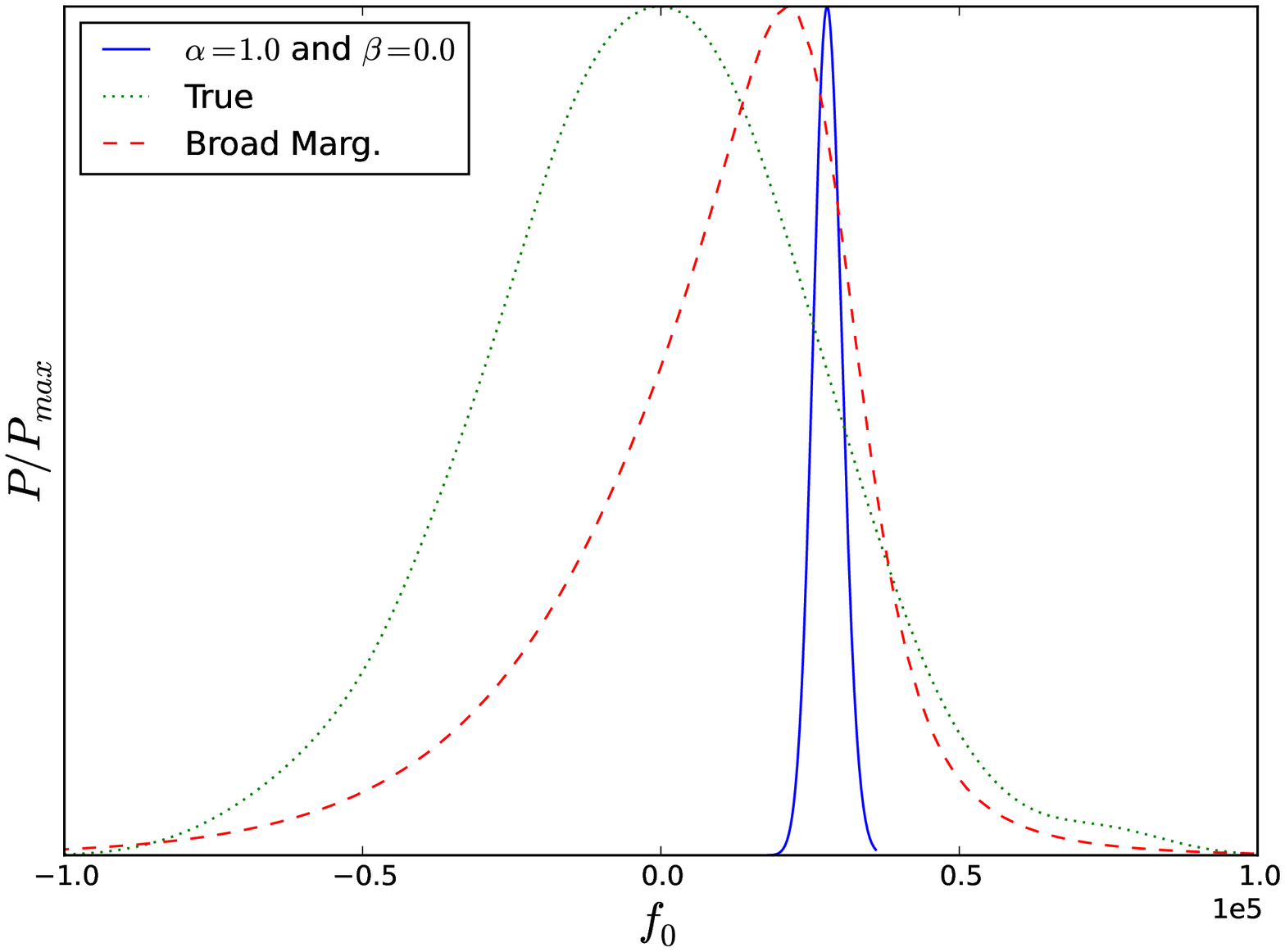}
\includegraphics[width=0.3\textwidth]{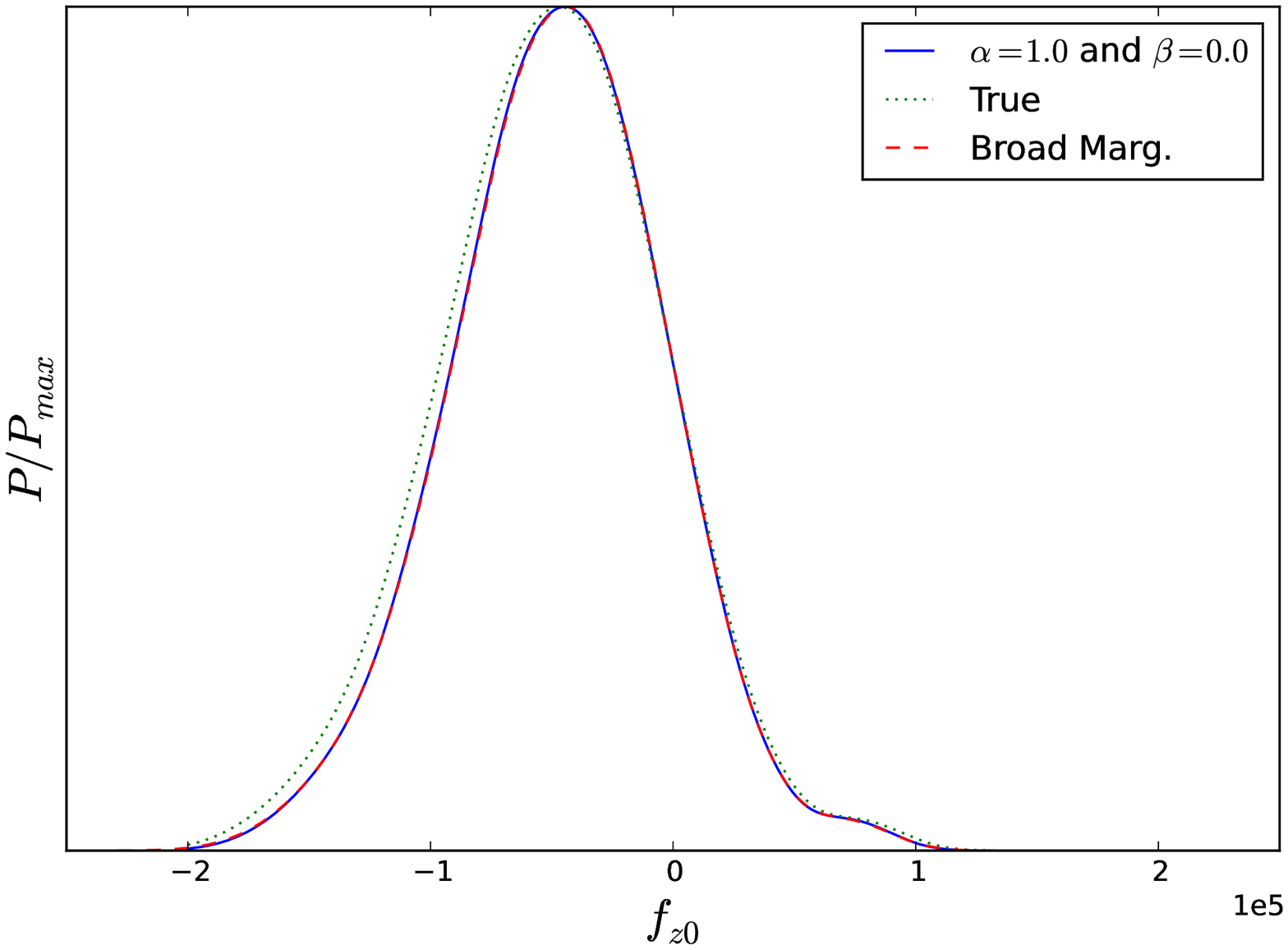}
\includegraphics[width=0.3\textwidth]{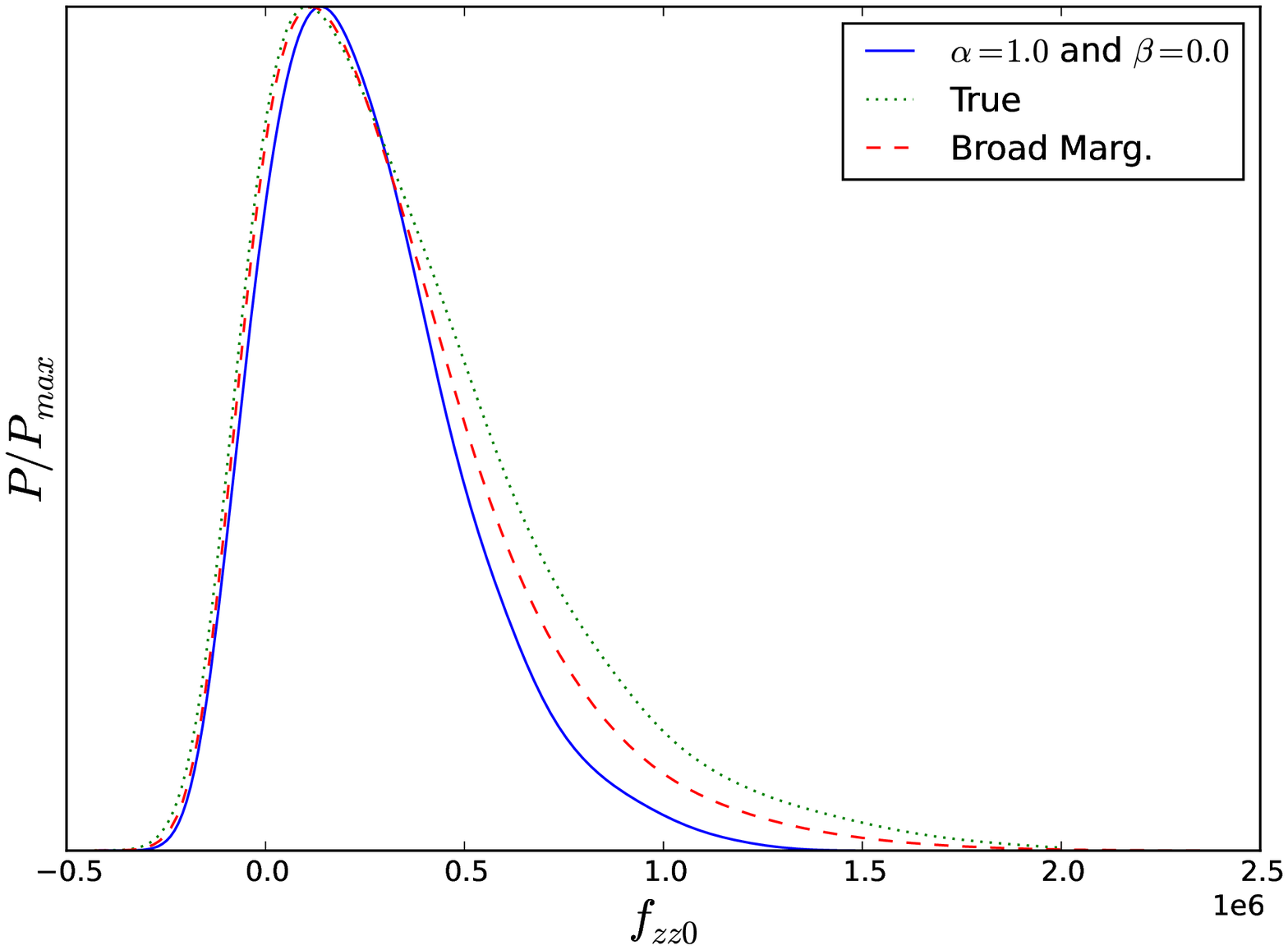}
\end{center}
\caption{Effects of the marginalization for different choices of the free parameters $\alpha$ and $\beta$ for $f(R)$ theories.}
\label{fig3}
\end{figure*}

As an example, where higher-order derivatives are involved, let us now consider the case of $f(R)$ gravity:
\begin{eqnarray}
\label{lagr f(R)}
\mathcal{S}\,=\,\int \text{d}^4 x \sqrt{-g}\left[\frac{1}{2} f(R)+{\cal L}_{m}\right]\,.
\end{eqnarray}
The $f(R)$ function and its derivatives with respect to the redshift $z$ evaluated today lead to
\begin{eqnarray}
&&\frac{f_0}{6H_0^2}\,=\,-\alpha q_0 +\Omega_m + 6\beta\left( 2+q_0-j_0\right)\ , \nn
&&\frac{f_{z0}}{6H_0^2}\,=\,\alpha\left(2+q_0-j_0\right)\ , \label{f2z0} \\
&&\frac{f_{2z0}}{6H_0^2}\,=\,6\beta\left(2+q_0-j_0\right)^2+\alpha\left[2+4q_0+(2+q_0)j_0+s_0\right]\,.\nonumber 
\end{eqnarray}
In this case, there are two extra free parameters in comparison with quintessence, namely
\begin{eqnarray}
\frac{{\rm d}f}{{\rm d}R}\biggr\rvert_{R=R_0}=\alpha\;\;\;;\;\;\; \frac{{\rm d}^2f}{{\rm d}R^2}\biggr\rvert_{R=R_0}=\frac{\beta}{H_0^2}\,.
\label{Conditions_1}
\end{eqnarray}
Previous works in the literature \cite{Capozziello_PRD_fR} fixed the values of $\alpha=1$ and $\beta=0$ a priori, such that the model coincides with General Relativity at $z=0$.
However, whenever  ${\rm d}^2f/{\rm d}R^2=0$ either a singularity or instability occurs \cite{Pogosian:2007sw}. Apart from this theoretical shortcoming, it turns out that cosmological values  $\alpha\neq 1$ and $\beta\neq 0$  may still produce viable cosmological models. Therefore, the naive assumption about the $\alpha$ and $\beta$  parameters must be abandoned and these  two parameters should consequently enter in the analysis as free parameters. It follows therefore, that it is not possible anymore to have one-to-one correspondence between the $f(R)$-derivatives and the cosmographic parameters. This means one should constrain the cosmographic parameters and use eqs. (\ref{f2z0}) and  (\ref{Conditions_1})  in order to reconstruct $f$ and its derivatives. As one can see, the data does not provide any constraints over $\alpha$ and $\beta$, so either priors over these parameters or complementary tests are 
necessary \cite{Refs_fR_tests}. In order to illustrate the difficulties in defining sensible priors over the aforementioned parameters, we generated mock data for the following toy-model: $f(R)=R+aR^2+bR^3$ with $\alpha=2.81$ and $\beta=0.06$. In Fig.~\ref{fig3}, we show the probability for $\{f_0, f_{z0}, f_{zz0}\}$ after fitting the model with the generated data. We have assumed three different hypotheses: the true values of $\{\alpha, \beta\}$, $\{\alpha=1, \beta=0\}$ and a ``broad'' marginalisation ($\alpha \sim N(1,0.05)$ and $\beta \sim N(0.07,0.05)$). As shown in Fig.~\ref{fig3}, the probability of $f_0$ is highly dependent on the choice of $\{\alpha, \beta\}$ which may even lead to ruling out the true values of $f_0$, when they are not known in advance - as is the case when dealing with real data. The differences for the values of $f_{z0}$ and $f_{zz0}$ are less prominent. Note that the values for the ``broad'' marginalisation do not cover the true values. However, the errors are very large for every case, which leads to a completely degenerated fit, such that a wide range of completely different $f(R)$ models lie in the 1$\sigma$ region. For example, by considering the viable Hu-Sawicki model \cite{HuSawicki}, it is straightforward to show that the $\{f_0, f_{z0}, f_{zz0}\}$ of this model, for a wide range of random values of its free parameters, lie within the $1\sigma$ region in Fig.~\ref{fig3}, despite the fact that the mock data was generated by a very different model. Consequently cosmography is not able to distinguish between them.  Hence, if one decides to marginalise considering all possible values for $\alpha$ and $\beta$, no useful constraints can be derived. 

On the other hand, priors can bias the results strongly, which may be thought of as an avoidable limitation for current data given the size of errors, but may become an issue when hundreds of thousands of supernovae have been observed. Therefore, we see that the cosmographic approach is extremely weak at reconstructing $f(R)$ gravity theories and therefore unable to rule out radically different kinds of $f(R)$ theories.
The same issue arises when dealing with theories which have higher-order derivatives in the matter sector. In order to show this, let us just obtain the expansions within the simplest Galileon model \cite{Deffayet:2009mn}:
\begin{eqnarray}
\label{lagrGalileon}
\mathcal{S}\,=&&\,\int \text{d}^4 x \sqrt{-g}\left[\frac{1}{2} R+\frac{1}{2}\left(c_1\phi+c_2\partial_{\mu}\phi\partial^{\mu}\phi\right.\right.\nn
&&\left.\left.+\,c_3\partial_{\mu}\phi\partial^{\mu}\phi\Box\phi\right)+{\cal L}_{m}\right]\,,
\end{eqnarray}
where  $c_{i=1,2,3}$ are coupling constants. The field can be expanded as
\bea
&&\frac{c_1}{6H_0^2}\phi_0=-1-\alpha_{G}-\beta_{G}+\Omega_m\ , \nn
&&\frac{\phi_{zz0}}{2\phi_{z0}}=-\frac{1}{\beta_G}+q_0-3\frac{\alpha_G}{\beta_G}-2\left[1+\frac{1}{4}(1+q_0)\right],
\label{G3}
\eea
where $\alpha_G\equiv c_2\phi_{0z}^{2}/6$ and $\beta_G\equiv c_3H_0^2\phi_{0z}^3$. 
Thus, the same issues as in $f(R)$ gravity appear, due to two additional free parameters $\alpha_G$ and $\beta_G$. The only advantage Galileons theories have relate to the errors which are not as large as in the $f(R)$ gravity case, since the gravitational sector does not include higher-order derivatives. Consequently the expansion of the  scalar field up to second order only depends on $q_0$. In conclusion, the reconstruction process based on cosmography for theories containing extra degrees of freedom -- not only $f(R)$ gravities or Galileons but also other extended theories as for instance,  Horndeski-like Lagrangians - leads to completely unconstrained parameters of the models under consideration.   
\section{Conclusions}
In this paper, we investigated several issues that emerge when using the cosmographic approach to constrain or reconstruct cosmological models, which have been overlooked in the previous literature. We first compared the cosmographic approach using different independent variables and found that when using the so-called $y$-redshift, biased constrains and larger errors are found on the cosmographic parameters compared to usual redshift $z$, even though $y$ appears to be theoretically better motivated.  

We then looked at what would happen if we used a fiducial model slightly different from $\Lambda$CDM, expecting that the fit of the cosmographic parameters would show a clear deviation away from the $j_0$ value in $\Lambda$CDM $(j_0=1)$. Instead we found that the results depend considerably on the order of the cosmographic expansion, leading to ambiguous constraints (as shown in Fig.~\ref{fig2}).

Finally, and motivated by previous works in the literature, we attempted to use cosmographic methods to reconstruct different models of dark energy.  We found that when considering theories with no higher-order derivatives in the gravitational or matter sectors, the method gives a clear picture of the underlying model, even though the errors are usually larger than those obtained using other approaches. However, when dealing with theories whose Lagrangians contain higher-order derivatives in either the gravitational or matter sector, there are extra free parameters which cannot be constrained by cosmography and must be marginalised in order to recover the underlying theory. As we illustrated, this fact leads in general to large errors preventing us from ruling models out with very different cosmological background evolution. 
Therefore, given the current state of affairs, it seems that the cosmographic approach is not a useful tool for theory reconstruction.

The focus here has been to highlight a number of shortcomings of the cosmographic approach, but we should also ask if there are other limitations that need to be taken into account and whether it is possible to overcome the limitations discussed in this paper.
Regarding the first question - there are two effects which were not considered in our analysis. The first one relates to the role of spatial curvature in the constraints, since it is known that $\Omega_k \neq 0$ can induce a time variation for the dark energy equation of state \cite{chris2007}.
Also, effects such as gravitational and Doppler lensing \cite{grav_lens}, or even local gravitational redshifts \cite{wojtak} lead to an extra scatter in the Hubble diagram, which degrades cosmological constraints when using SNe Ia data. Although quantified in a $\Lambda$CDM model using simulations and inserted as an extra error in SNe Ia analysis \cite{union2.1}, the impact on extended gravity theories may be more important and a thorough study of this effect is necessary in order to rule out those theories for which the cosmographic approach cannot be applied.

Finally in response to the second question - we can identify a number of challenges faced by this method: There needs to be $i)$ a clear definition of auxiliary variables, together with their range of applicability and extensive testing against mock data; $ii)$ a robust statistical method which could in principle establish a trade-off between number of data points, number of cosmographical parameters and Bayesian evidence in order to provide criteria to safely rule out models; and $iii)$ very well motivated priors over the extra parameters in higher-order theories in order to get useful constraints for these theories, even when neglecting lensing and local effects. Only when these issues are addressed, can we be confident that cosmography is a useful tool for constraining extended theories of gravity.

\begin{acknowledgments}
V.C.B. is supported by CNPq-Brazil through a fellowship within the program Science without Borders. 
A.d.l.C.D. acknowledges financial support from University of Cape Town Launching Grant Programme and MINECO (Spain) projects 
FIS2014-52837-P, FPA2014-53375-C2-1-P and Consolider-Ingenio MULTIDARK CSD2009-00064.
%
%
%
P. K. S. D. thanks the NRF for financial support.
D.S.-G. acknowledges support from a postdoctoral fellowship Ref.~SFRH/BPD/95939/2013 by Funda\c{c}\~ao para a Ci\^encia e a Tecnologia (FCT, Portugal) and the support through the research grant UID/FIS/04434/2013 (FCT, Portugal). D.S.-G. also acknowledges the NRF financial support from the University of Cape Town (South Africa). 
We would like to thank the referees for providing useful comments which have served to improve the final version of the paper.

\end{acknowledgments}

\end{document}